\documentclass[preprint,aps,showpacs,nofootinbib,preprintnumbers,amsmath,amssymb]{revtex4-1}
\usepackage{epsfig}
\usepackage{subfigure}
\usepackage{dcolumn}
\usepackage{bm}
\usepackage[usenames ,dvipsnames]{xcolor}
\usepackage{slashed}
\usepackage{graphicx,color}
\usepackage{amsmath}
\usepackage{nonfloat}
\begin{document}
\title{Three-body charmed baryon Decays with SU(3) flavor symmetry}

\author{C.Q. Geng$^{1,2,3}$, Y.K. Hsiao$^{1}$, Chia-Wei Liu$^{2}$ and Tien-Hsueh Tsai$^{2}$}
\affiliation{
	$^{1}$School of Physics and Information Engineering, Shanxi Normal University, Linfen 041004\\
	$^{2}$Department of Physics, National Tsing Hua University, Hsinchu 300\\
	$^{3}$Physics Division, National Center for Theoretical Sciences, Hsinchu 300
}
\date{\today}
\begin{abstract}
We study the three-body anti-triplet ${\bf B_c}\to {\bf B_n}MM'$ decays with the $SU(3)$ flavor ($SU(3)_f$) symmetry, where ${\bf B_c}$ denotes the charmed baryon anti-triplet of $(\Xi_c^0,-\Xi_c^+,\Lambda_c^+)$, and  ${\bf B_n}$ and $M(M')$ represent baryon and meson octets, respectively.  
 By considering only the S-wave $MM'$-pair contributions without resonance effects, the decays of ${\bf B_c}\to {\bf B_n}MM'$ can be decomposed into irreducible forms with 11 parameters under $SU(3)_f$, which are fitted by the 14 existing data, resulting in a reasonable value of $\chi^2/d.o.f=2.8$ for the fit.  Consequently, we find that the triangle sum rule of  ${\cal A}(\Lambda_c^+\to n\bar K^0 \pi^+)-{\cal A}(\Lambda_c^+\to pK^- \pi^+)-\sqrt 2 {\cal A}(\Lambda_c^+\to p\bar K^0 \pi^0)=0$ given by the isospin symmetry holds
  under $SU(3)_f$,  where ${\cal A}$ stands for the decay amplitude. In addition, we predict that ${\cal B}(\Lambda_c^+\to n \pi^{+} \bar{K}^{0})=(0.9\pm 0.8)\times 10^{-2}$, which is  $3-4$ times smaller than the BESIII observation, indicating the existence of the resonant states. For the to-be-observed ${\bf B_c}\to {\bf B_n}MM'$ decays, we compute the branching fractions with the $SU(3)_f$ amplitudes to be compared to the  BESIII and LHCb measurements in the future.
\end{abstract}
\maketitle

\section{introduction}
The three-body charmed baryon ${\bf B_c}\to {\bf B_n} MM'$ decays  
have been recently searched  by the experimental  Collaborations of   BELLE, BESIII and LHCb,
where ${\bf B_c}\equiv (\Xi_c^0,-\Xi_c^+,\Lambda_c^+)$ 
denotes the charmed baryon anti-triplet,
and  $\bf B_n$ and $M^{(')}$ correspond to 
the baryon and meson octets, respectively.
For example, the decay of $\Lambda_c^+\to p K^-\pi^+$ 
has  been observed 
with high precision by BELLE and BESIII~\cite{Zupanc:2013iki,Ablikim:2015flg},
which improves the accuracy of the $\Lambda_b$ decays
with $\Lambda_c^+$ as one of the final states~\cite{pdg}.
Besides, the crucial information on the higher wave baryon resonances
like $\Lambda(1405)$ has been extracted from 
the $\Sigma\pi$ invariant mass spectra
of the $\Lambda_c^+\to \Sigma\pi\pi$ decays~\cite{Berger:2018pli}.
The another interest comes from the test of the theoretical approach.
For example, the first observation of 
$\Lambda_c^+\to n K^0_s \pi^+$ has been used to
examine the isospin relation~\cite{Ablikim:2016mcr}, that is,
$R(\Delta)\equiv {\cal A}(\Lambda_c^+\to n\bar K^0 \pi^+)+
{\cal A}(\Lambda_c^+\to pK^- \pi^+)
+\sqrt 2{\cal A}(\Lambda_c^+\to p\bar K^0 \pi^0)=0$~\cite{Lu:2016ogy,Gronau:2018vei}\footnote{To calculate the decay amplitude
of ${\cal A}$, we use the conventions of $| \pi^{+} \rangle = -| 1 1 \rangle$ and $| \bar{K}^0 \rangle = -| \frac{1}{2} \frac{1}{2} \rangle$, 
  whereas 
  $| \pi^{+} \rangle = |1 1 \rangle$ and $| \bar{K}^{0} \rangle = | \frac{1}{2} \frac{1}{2} \rangle$  
  are taken in Refs.~\cite{Lu:2016ogy,Gronau:2018vei}, resulting in
the relation to be $R(\Delta)\equiv {\cal A}(\Lambda_c^+\to n\bar K^0 \pi^+)-{\cal A}(\Lambda_c^+\to pK^- \pi^+)
-\sqrt 2{\cal A}(\Lambda_c^+\to p\bar K^0 \pi^0)=0$. 
However, the different signs in $R(\Delta)$ and other similar relations do not affect the physical consequences of these relations due to
the arbitrariness of the phase of the amplitude.}

Since 
the $\Lambda_c^+\to p K^- \pi^+$ decay
shares the similar diagrams as
the doubly Cabibbo-suppressed $\Lambda_c^+\to p K^+ \pi^-$ one, we have
the  ratio of ${\cal B}(\Lambda_c^+\to p K^+ \pi^-)/{\cal B}(\Lambda_c^+\to p K^- \pi^+)
={\cal R}_{K\pi}\tan^4\theta_c$
with ${\cal R}_{K\pi}\simeq 1.0$ and $\theta_c$ the Cabibbo angle should hold.
Nonetheless, the values of ${\cal R}_{K\pi}=0.82\pm 0.12$~\cite{Yang:2015ytm} and $0.58\pm 0.06$~\cite{Aaij:2017rin} have been measured by BELLE and  LHCb, respectively, showing a possible deviation caused by an additional 
$W$-exchange amplitude for $\Lambda_c^+\to p K^-\pi^+$.
As a result, the  ${\bf B_c}\to {\bf B_n} MM'$ decays are important 
for achieving a deeper insight into the hadronization of particle interactions.

In contrast with the abundant observations, there rarely exist 
systematic  theoretical studies on the ${\bf B_c}\to {\bf B_n} MM'$ decays,
apart from those based on the isospin symmetry~\cite{Lu:2016ogy,Gronau:2018vei}.
This is due to the fact that the scale of the charm quark mass ($m_c$) 
is too large for the flavor $SU(3)$  ($SU(3)_f$)  symmetry, but
the theories based on  the heavy quark expansion may  not be valid as $m_c$ is not large enough.
In addition,
the factorization fails to work well 
in the charmed hadron decays~\cite{Bjorken:1988ya,Lu:2016ogy},
whereas it is successfully  used 
in the beauty hadron ones~\cite{ali,Geng:2006jt,Hsiao:2014mua}.
The alternative approaches for the charmed hadron decays 
have been shown in Refs.~~\cite{Cheng:1991sn,Cheng:1993gf,
Zenczykowski:1993hw,Fayyazuddin:1996iy,Dhir:2015tja,Cheng:2018hwl},
which take into account the non-factorizable effects.
On the other hand, 
the $SU(3)_f$ symmetry 
has been tested as a useful tool both
in the beauty and charmed hadron decays~\cite{He:2000ys,Fu:2003fy,Hsiao:2015iiu,
He:2015fwa,He:2015fsa,Cheng:2012xb,Pirtskhalava:2011va,Grossman:2012ry},
particularly, 
the two-body ${\bf B_c}\to {\bf B_n} M$ decays~\cite{Savage:1989qr,
Savage:1991wu,h_term,Lu:2016ogy,Geng:2017esc,Geng:2018plk,Wang:2017gxe,
Wang:2017azm,Geng:2017mxn,Geng:2018bow,Geng:2018rse}.
It is hence expected that the same symmetry can be applied to  
the three-body ${\bf B_c}\to {\bf B_n} MM'$ decays.
In this paper, 
we will relate
the possible ${\bf B_c}\to {\bf B_n} MM'$ decay processes 
with the $SU(3)_f$ parameters~\cite{Savage:1989qr}, by which
the systematic numerical analysis can be performed for the first time.
Under the $SU(3)_f$ symmetry,
we will derive the relation of ${\cal R}(\Delta)=0$, and examine the value of 
 ${\cal R}_{K\pi}$ from the ratio of 
${\cal B}(\Lambda_c^+\to p K^+ \pi^-)/{\cal B}(\Lambda_c^+\to p K^- \pi^+)$.

Our paper is organized as follows.
We give the formalism in  Sec. II,
where the amplitudes for the three-body charmed baryon decays under the $SU(3)_f$ symmetry
are presented.
 In Sec. III, we show our numerical results  and discussions.
 Our conclusions are  in Sec. IV.

\section{Formalism}
The three-body ${\bf B_c}\to {\bf B_n} MM'$ decays
can proceed through the charm quark decays of
$c\to s u\bar d$, $c\to u d\bar d\,(u s\bar s)$ and $c\to d u\bar s$,
where ${\bf B_{c,n}}$ 
and $M^{(')}$ denote 
the  baryon 
and meson states, respectively. 
Accordingly, the tree-level effective Hamiltonian is given by~\cite{Buras:1998raa}
\begin{eqnarray}\label{Heff}
{\cal H}_{eff}&=&\sum_{i=-,+}\frac{G_F}{\sqrt 2}c_i
\bigg[V_{cs}V_{ud} O_i +V_{cd}V_{ud}O_i^{\dagger} +V_{cd}V_{us}O_i^{\prime}\bigg]\,,
\end{eqnarray}
where $G_F$ is the Fermi constant,
$c_{\pm}$ represent  the 
Wilson coefficients and
$V_{ij}$ correspond to the CKM matrix elements,
while $O_\mp$, $O_\mp^\dagger$ and $O_\mp^{\prime}$
are the four-quark operators, 
written as
\begin{eqnarray}\label{O12}
&&
O_\mp={1\over 2}\left[(\bar u d)(\bar s c)\mp (\bar s d)(\bar u c)\right]\,,\;\nonumber\\
&&
O^{\dagger}_\mp=
{1\over 2}\left[(\bar u d)(\bar d c)\mp (\bar d d)(\bar u c)\right]-
{1\over 2}\left[(\bar u s)(\bar s c)\mp (\bar s s)(\bar u c)\right]\,,\;\nonumber\\
&&
O'_\mp={1\over 2}\left[(\bar u s)(\bar d c)\mp (\bar d s)(\bar u c)\right]\,,
\end{eqnarray}
with $(\bar q_1 q_2)(\bar q_3 c)\equiv
\bar q_1\gamma_\mu(1-\gamma_5)q_2\,\bar q_3\gamma^\mu(1-\gamma_5)c$.
Here, the relation of  $V_{cs}V_{us}=-V_{cd}V_{ud}$ has been used for $O^{\dagger}_\mp$
to combine the $c\to u d\bar d$ and $c\to u s\bar s$ transitions. 
By means of the Cabibbo angle $\theta_c$,
it is given that 
$(V_{cs}V_{ud}$, $V_{cd}V_{ud}$, $V_{cd}V_{us})$
=$c^2_c(1,-t_c,-t_c^2)$,
where $(c_c,t_c)\equiv(\cos\theta_c,\tan\theta_c)$, 
such that the decays with
$O_\mp$, $O_\mp^\dagger$ and $O_\mp^{\prime}$ are classified as
the Cabibbo-favored (CF), Cabibbo-suppressed (CS), and
doubly Cabibbo-suppressed (DCS) processes, respectively.

In Eq.~(\ref{O12}), $(\bar q_1 q_2)(\bar q_3 c)$
can be rewritten as $(\bar q^i q_k \bar q^j)c$ 
with $q_i=(u,d,s)$ the triplet of 3 under the $SU(3)_f$ symmetry, 
by suppressing the Dirac and Lorentz indices.
Furthermore, 
since $(\bar q^i q_k \bar q^j)c$ can be decomposed as
the irreducible forms of
$(\bar 3\times 3\times \bar 3)c=(\bar 3+\bar 3'+6+\overline{15})c$, 
one derives that~\cite{Savage:1989qr}
\begin{eqnarray}\label{O_su3}
O_{-(+)}\simeq 
&{\cal O}_{6(\overline{15})}&
=\frac{1}{2}(\bar u d\bar s\mp\bar s d\bar u)c\,,\nonumber\\
O_{-(+)}^\dagger\simeq &{\cal O}_{6(\overline{15})}^\dagger&
=\frac{1}{2}(\bar u d\bar d\mp\bar d d\bar u)c
-\frac{1}{2}(\bar u s\bar s\mp\bar s s\bar u)c\,,\nonumber\\
O'_{-(+)}\simeq 
&{\cal O}'_{6(\overline{15})}&
=\frac{1}{2}(\bar u s\bar d\mp\bar d s\bar u)c\,.
\end{eqnarray}
Subsequently,
the effective Hamiltonian in Eq.~(\ref{Heff})
has the expression under the $SU(3)_f$ symmetry, 
given by~\cite{Geng:2017esc,Geng:2018plk,Geng:2017mxn,Geng:2018bow}
\begin{eqnarray}\label{Heff2}
{\cal H}_{eff}&=&\frac{G_F}{\sqrt 2} c_c^2 
\left[c_-  \frac{\epsilon^{ijl}}{2}H(6)_{lk}+c_+H(\overline{15})_k^{ij}\right]c\,,
\end{eqnarray}
where $H(6,\overline{15})$ are presented as the tensor forms of
$({\cal O}_{6}^{(\dagger,\prime)},{\cal O}_{\overline{15}}^{(\dagger,\prime)})$
in Eq.~(\ref{O_su3}). Their non-zero entries are 
given by~\cite{Savage:1989qr,Savage:1991wu}
\begin{eqnarray}\label{nz_entry}
&&H_{22}(6)=2\,,H_2^{13}(\overline{15})=H_2^{31}(\overline{15})=1\,,\nonumber\\
&&H_{23}(6)=H_{32}(6)=-2t_c\,,
H^2_{12}(\overline{15})=H^2_{21}(\overline{15})=t_c\,,\nonumber\\
&&H_{33}(6)=2t_c^2\,,
{H}_3^{12}(\overline{15})={H}_{3}^{21}(\overline{15})=-t_c^2\,,
\end{eqnarray}
with $(i,j,k)$ for the quark indices. 
Correspondingly, the three lowest-lying charmed baryon states of ${\bf B}_c$
form an anti-triplet of $\bar 3$ to consist of
$(ds-sd)c$, $(us-su)c$ and $(ud-du)c$,
and ${\bf B}_n(M)$ belongs to the baryon (meson) octet of 8, which are written as
\begin{eqnarray}\label{BM_SU3}
{\bf B}_{c}&=&(\Xi_c^0,-\Xi_c^+,\Lambda_c^+)\,,\nonumber\\
{\bf B}_n&=&\left(\begin{array}{ccc}
\frac{1}{\sqrt{6}}\Lambda^0+\frac{1}{\sqrt{2}}\Sigma^0 & \Sigma^- & \Xi^-\\
 \Sigma^+ &\frac{1}{\sqrt{6}}\Lambda^0 -\frac{1}{\sqrt{2}}\Sigma^0  & \Xi^0\\
 p & n &-\sqrt{\frac{2}{3}}\Lambda^0
\end{array}\right)\,,\nonumber\\
M&=&\left(\begin{array}{ccc} 
	\frac{1}{\sqrt{2}}\pi^0+\frac{1}{\sqrt{6}}\eta & \pi^- & K^-\\
	\pi^+ &-\frac{1}{\sqrt{2}}\pi^0+\frac{1}{\sqrt{6}}\eta& \bar K^0\\
	K^+ & K^0& -\sqrt{\frac{2}{3}}\eta
\end{array}\right)\,,
\end{eqnarray}
respectively.

Now, one is able to connect
the octets of $({\bf B}_n, M)^i_j$ and anti-triplet of $({\bf B}_c)_i$
to $(\epsilon^{ijl}H(6)_{lk},H(\overline{15})_k^{ij})$
in ${\cal H}_{eff}$ of Eq.~(\ref{Heff2}) to get the $SU(3)_f$ amplitudes.
Since the Wilson coefficients are scale-dependent,
in the NDR scheme it is calculated that $(c_-,c_+)=(1.78,0.76)$ 
at the scale $\mu=1$ GeV~\cite{Fajfer:2002gp,Li:2012cfa}.
The value of $(c_-/c_+)^2\simeq 5.5$ implies 
the suppressed branching ratios associated with $H(\overline{15})$.
Hence, we follow Refs.~\cite{Lu:2016ogy,Geng:2017esc,Wang:2017gxe} 
to ignore the amplitudes from $H(\overline{15})$.
By means of ${\cal A}({\bf B}_c\to {\bf B}_n MM')\equiv (G_F/\sqrt 2) T({\bf B}_c\to {\bf B}_n MM')$,
the T-amplitude of ${\bf B}_c\to {\bf B}_n MM'$ 
can be derived as~\cite{Savage:1989qr}
\begin{eqnarray}\label{3b_amp}
T({\bf B}_{c}\to {\bf B}_n MM)&=& 
a_1(\bar {\bf B}_n)^k_i (M)^m_l(M')^l_mH(6)_{jk}T^{ij}+a_2(\bar {\bf B}_n)^k_i (M)^m_j(M')^l_mH(6)_{kl}T^{ij}\nonumber\\
&+&
a_3(\bar {\bf B}_n)^k_i (M)^m_k(M')^l_mH(6)_{jl}T^{ij}+a_4(\bar {\bf B}_n)^k_i (M)^l_j(M')^m_kH(6)_{lm}T^{ij}\nonumber\\
&+&
a_5(\bar {\bf B}_n)^l_k (M)^m_j(M')^k_mH(6)_{il}T^{ij}+a_6(\bar {\bf B}_n)^l_k (M)^m_j(M')^k_lH(6)_{im}T^{ij}\,,
%
\end{eqnarray}
with $T^{ij}=({\bf B_c})_a\epsilon^{aij}$,
where $c_c^2$ and $c_-$ in ${\cal H}_{eff}$ have been absorbed into 
the $SU(3)_f$ parameters $a_i$ ($i=1, 2, ..., 6$).
While there exists 
the relative orbital angular momentum $L$
between the two-meson states, 
we have assumed the S-wave $MM'$-pair ($L=0$)
in the dominant amplitudes in Eq.~(\ref{3b_amp}),
whereas the P-wave one ($L=1$) is neglected.
However, 
there are some cases in which the S-wave contributions vanish, 
but P-wave ones  are  dominant, resulting in 
the another set of amplitudes to be studied elsewhere.
For example, the decay of 
$\Lambda_c^+\to \Lambda\pi^+\pi^0$
with the measured branching ratio around $7.1\%$ 
is mainly from the P-wave contribution. 

The integration over the phase space of the three-body decay
relies on the equation of~\cite{pdg}
\begin{eqnarray}
\Gamma=\int_{m_{12}^2}\int_{m_{23}^2}
\frac{1}{(2\pi)^3}\frac{|{\cal A}({\bf B}_c\to {\bf B}_n MM')|^2}{32m^3_{\bf B_c}}
dm_{12}^2 dm_{23}^2\,,
\end{eqnarray}
where $m_{12}=p_M+p_{M'}$, $m_{23}=p_{M'}+p_{\bf B_n}$
and ${\cal A}({\bf B}_c\to {\bf B}_n MM')$ is related to $T({\bf B}_c\to {\bf B}_n MM')$ given in Eq.~(\ref{3b_amp}).
In Tables~\ref{Tamp1}, \ref{Tamp2} and \ref{Tamp3}, we show the full expansions of 
$T(\Lambda_c^+\to {\bf B}_n MM')$, $T (\Xi_{c}^{+} \to {\bf B}_n MM')$ and
$T(\Xi_{c}^{0} \to {\bf B}_n MM')$,
respectively.
In general, the  $SU(3)_f$ parameters depend on  $m_{12}$ and $m_{23}$. 
However,  all structures in the Dalitz plots come from the dynamical effects, such as those from
the  resonant states. 
Clearly, the squared amplitude in the Dalitz plot is almost structureless for the decay without the resonance. 
As a result,  we treat our decay amplitudes as constants without energy dependences so that  they can be factored out from the integrals 
as an approximation.

\begin{table}
\caption{T-amplitudes of $ \Lambda_{c}^{+}  \to {\bf B_n}MM'$.}\label{Tamp1}
{
\scriptsize
\begin{tabular}{|c|c|}
\hline
CF mode
& T-amp\\
\hline
$ \Sigma^{+} \pi^{0} \pi^{0} $ & $ 4a_{1} + 2a_{2} + 2a_{3} + 2a_{4} - 2a_{5} $ \\
$ \Sigma^{+} \pi^{+} \pi^{-} $ & $ 4a_{1} + 2a_{2} + 2a_{3} - 2a_{5} - 2a_{6} $ \\
			$ \Sigma^{+} K^{0} \bar{K}^{0} $ & $ 4a_{1} + 2a_{2} + 2a_{3} $ \\
			$ \Sigma^{+} K^{+} K^{-} $ & $ 4a_{1} - 2a_{5} $ \\
			$ \Sigma^{+} \eta^{0} \eta^{0} $ & $ 4a_{1} +\frac{ 2a_{2}}{3}+\frac{ 2a_{3}}{3}+\frac{ 2a_{4}}{3}-\frac{ 2a_{5}}{3} $ \\
			$ \Sigma^{0} \pi^{0} \pi^{+} $ & $ -2a_{4} - 2a_{6} $ \\
			$ \Sigma^{0} K^{+} \bar{K}^{0} $ & $ \sqrt{2}a_{2} + \sqrt{2}a_{3} + \sqrt{2}a_{5} $ \\
			$ \Sigma^{-} \pi^{+} \pi^{+} $ & $ -4a_{4} - 4a_{6} $ \\
			$ \Xi^{0} \pi^{0} K^{+} $ & $ -\sqrt{2}a_{5} $ \\
			$ \Xi^{0} \pi^{+} K^{0} $ & $ -2a_{5} - 2a_{6} $ \\
			$ \Xi^{-} \pi^{+} K^{+} $ & $ -2a_{6} $ \\
			$ p \pi^{0} \bar{K}^{0} $ & $ -\sqrt{2}a_{3} - \sqrt{2}a_{4} $ \\
			$ p \pi^{+} K^{-} $ & $ 2a_{3} - 2a_{6} $ \\
			$ p \bar{K}^{0} \eta^{0} $ & $ -\frac{\sqrt{6}a_{3}}{3}+\frac{ \sqrt{6}a_{4}}{3} $ \\
			$ n \pi^{+} \bar{K}^{0} $ & $ -2a_{4} - 2a_{6} $ \\
			$ \Lambda^{0} \pi^{+} \eta^{0} $ & $ -\frac{2a_{2}}{3}+\frac{ 2a_{3}}{3}-\frac{ 2a_{5}}{3}- 2a_{6} $ \\
			$ \Lambda^{0} K^{+} \bar{K}^{0} $ & $ -\frac{\sqrt{6}a_{2}}{3}+\frac{ \sqrt{6}a_{3}}{3}-\frac{ \sqrt{6}a_{5}}{3} $ \\
			\hline
\end{tabular}
\begin{tabular}{|c|c|}
			\hline
                         CS mode
			& T-amp/$t_c$\\

			\hline
			$ \Sigma^{+} \pi^{0} K^{0} $ & $ \sqrt{2}a_{2} + \sqrt{2}a_{3} + 2\sqrt{2}a_{4} $ \\
			$ \Sigma^{+} \pi^{-} K^{+} $ & $ -2a_{2} - 2a_{3} + 2a_{6} $ \\
			$ \Sigma^{+} K^{0} \eta^{0} $ & $ \frac{\sqrt{6}a_{2}}{3}+\frac{ \sqrt{6}a_{3}}{3}-\frac{ 2\sqrt{6}a_{4}}{3} $ \\
			$ \Sigma^{0} \pi^{+} K^{0} $ & $ -\sqrt{2}a_{2} - \sqrt{2}a_{3} - 2\sqrt{2}a_{4} $ \\
			$ \Sigma^{0} K^{+} \eta^{0} $ & $ \frac{\sqrt{3}a_{2}}{3}+\frac{ \sqrt{3}a_{3}}{3}-\frac{ 2\sqrt{3}a_{4}}{3} $ \\
			$ \Sigma^{-} \pi^{+} K^{+} $ & $ 4a_{4} + 2a_{6} $ \\
			$ p \pi^{0} \pi^{0} $ & $ -4a_{1} - 2a_{2} + 2a_{5} $ \\
			$ p \pi^{0} \eta^{0} $ & $ \frac{2\sqrt{3}a_{2}}{3}-\frac{ 2\sqrt{3}a_{4}}{3}+\frac{ 2\sqrt{3}a_{5}}{3} $ \\
			$ p \pi^{+} \pi^{-} $ & $ -4a_{1} - 2a_{2} + 2a_{5} $ \\
			$ p K^{+} K^{-} $ & $ -4a_{1} - 2a_{3} + 2a_{5} + 2a_{6} $ \\
			$ p \eta^{0} \eta^{0} $ & $ -4a_{1} -\frac{ 2a_{2}}{3}-\frac{ 8a_{3}}{3}+\frac{ 4a_{4}}{3}+\frac{ 2a_{5}}{3} $ \\
			$ n \pi^{+} \eta^{0} $ & $ \frac{2\sqrt{6}a_{2}}{3}-\frac{ 2\sqrt{6}a_{4}}{3}+\frac{ 2\sqrt{6}a_{5}}{3} $ \\
			$ n K^{+} \bar{K}^{0} $ & $ 2a_{2} + 2a_{4} + 2a_{5} + 2a_{6} $ \\
			$ \Lambda^{0} \pi^{0} K^{+} $ & $ \frac{\sqrt{3}a_{2}}{3}-\frac{ \sqrt{3}a_{3}}{3}-\frac{ 2\sqrt{3}a_{5}}{3} $ \\
			$ \Lambda^{0} \pi^{+} K^{0} $ & $ \frac{\sqrt{6}a_{2}}{3}-\frac{ \sqrt{6}a_{3}}{3}-\frac{ 2\sqrt{6}a_{5}}{3} $ \\
			$ \Lambda^{0} K^{+} \eta^{0} $ & $ -\frac{a_{2}}{3}+\frac{ a_{3}}{3}+\frac{ 2a_{5}}{3}+ 2a_{6} $ \\
			&\\
			\hline
\end{tabular}
\begin{tabular}{|c|c|}
			\hline
			DCS mode
			& T-amp/$t_c^2$\\
			\hline
			$ \Sigma^{+} K^{0} K^{0} $ & $ 4a_{4} $ \\
			$ \Sigma^{0} K^{0} K^{+} $ & $ 2\sqrt{2}a_{4} $ \\
			$ \Sigma^{-} K^{+} K^{+} $ & $ -4a_{4} $ \\
			$ p \pi^{0} K^{0} $ & $ -\sqrt{2}a_{2} $ \\
			$ p \pi^{-} K^{+} $ & $ 2a_{2} $ \\
			$ p K^{0} \eta^{0} $ & $ -\frac{\sqrt{6}a_{2}}{3}-\frac{ 2\sqrt{6}a_{4}}{3} $ \\
			$ n \pi^{0} K^{+} $ & $ -\sqrt{2}a_{2} $ \\
			$ n \pi^{+} K^{0} $ & $ -2a_{2} $ \\
			$ n K^{+} \eta^{0} $ & $ \frac{\sqrt{6}a_{2}}{3}+\frac{ 2\sqrt{6}a_{4}}{3} $ \\	
			&\\
			&\\
			&\\
			&\\
			&\\
			&\\
			&\\
			&\\
			\hline
\end{tabular}
}
\end{table}

\begin{table}
\caption{T-amplitudes of $ \Xi_{c}^{+}  \to {\bf B_n}MM'$.}\label{Tamp2}
{
\scriptsize
\begin{tabular}{|c|c|}
			\hline
			CF mode& T-amp\\
			\hline
			$ \Sigma^{+} \pi^{0} \bar{K}^{0} $ & $ -\sqrt{2}a_{2} - \sqrt{2}a_{4} $ \\
			$ \Sigma^{+} \pi^{+} K^{-} $ & $ 2a_{2} $ \\
			$ \Sigma^{+} \bar{K}^{0} \eta^{0} $ & $ -\frac{\sqrt{6}a_{2}}{3}+\frac{ \sqrt{6}a_{4}}{3} $ \\
			$ \Sigma^{0} \pi^{+} \bar{K}^{0} $ & $ \sqrt{2}a_{4} $ \\
			$ \Xi^{0} \pi^{0} \pi^{+} $ & $ \sqrt{2}a_{4} $ \\
			$ \Xi^{0} \pi^{+} \eta^{0} $ & $ -\frac{2\sqrt{6}a_{2}}{3}-\frac{ \sqrt{6}a_{4}}{3} $ \\
			$ \Xi^{0} K^{+} \bar{K}^{0} $ & $ -2a_{2} $ \\
			$ \Xi^{-} \pi^{+} \pi^{+} $ & $ -4a_{4} $ \\
			$ p \bar{K}^{0} \bar{K}^{0} $ & $ 4a_{4} $ \\
			$ \Lambda^{0} \pi^{+} \bar{K}^{0} $ & $ \sqrt{6}a_{4} $ \\
			&\\
			&\\
			&\\
			&\\
			&\\
			&\\
			&\\
			&\\
			&\\
			\hline
\end{tabular}
\begin{tabular}{|c|c|}
			\hline
			CS mode& T-amp/$t_c$\\
			\hline
			$ \Sigma^{+} \pi^{0} \pi^{0} $ & $ -4a_{1} - 2a_{3} + 2a_{5} $ \\
			$ \Sigma^{+} \pi^{0} \eta^{0} $ & $ \frac{2\sqrt{3}a_{3}}{3}-\frac{ 2\sqrt{3}a_{4}}{3}+\frac{ 2\sqrt{3}a_{5}}{3} $ \\
			$ \Sigma^{+} \pi^{+} \pi^{-} $ & $ -4a_{1} - 2a_{3} + 2a_{5} + 2a_{6} $ \\
			$ \Sigma^{+} K^{+} K^{-} $ & $ -4a_{1} - 2a_{2} + 2a_{5} $ \\
			$ \Sigma^{+} \eta^{0} \eta^{0} $ & $ -4a_{1} -\frac{ 8a_{2}}{3}-\frac{ 2a_{3}}{3}+\frac{ 4a_{4}}{3}+\frac{ 2a_{5}}{3} $ \\
			$ \Sigma^{0} \pi^{0} \pi^{+} $ & $ 2a_{6} $ \\
			$ \Sigma^{0} \pi^{+} \eta^{0} $ & $ -\frac{2\sqrt{3}a_{3}}{3}+\frac{ 2\sqrt{3}a_{4}}{3}-\frac{ 2\sqrt{3}a_{5}}{3} $ \\
			$ \Sigma^{0} K^{+} \bar{K}^{0} $ & $ -\sqrt{2}a_{3} - \sqrt{2}a_{4} - \sqrt{2}a_{5} $ \\
			$ \Sigma^{-} \pi^{+} \pi^{+} $ & $ 4a_{6} $ \\
			$ \Xi^{0} \pi^{0} K^{+} $ & $ \sqrt{2}a_{2} - \sqrt{2}a_{4} + \sqrt{2}a_{5} $ \\
			$ \Xi^{0} \pi^{+} K^{0} $ & $ 2a_{2} + 2a_{4} + 2a_{5} + 2a_{6} $ \\
			$ \Xi^{0} K^{+} \eta^{0} $ & $ -\frac{\sqrt{6}a_{2}}{3}+\frac{ \sqrt{6}a_{4}}{3}-\frac{ \sqrt{6}a_{5}}{3} $ \\
			$ \Xi^{-} \pi^{+} K^{+} $ & $ 4a_{4} + 2a_{6} $ \\
			$ p \pi^{0} \bar{K}^{0} $ & $ \sqrt{2}a_{2} + \sqrt{2}a_{3} $ \\
			$ p \pi^{+} K^{-} $ & $ -2a_{2} - 2a_{3} + 2a_{6} $ \\
			$ p \bar{K}^{0} \eta^{0} $ & $ \frac{\sqrt{6}a_{2}}{3}+\frac{ \sqrt{6}a_{3}}{3}+\frac{ 4\sqrt{6}a_{4}}{3} $ \\
			$ n \pi^{+} \bar{K}^{0} $ & $ 2a_{6} $ \\
			$ \Lambda^{0} \pi^{+} \eta^{0} $ & $ -\frac{4a_{2}}{3}-\frac{ 2a_{3}}{3}+ 2a_{4} +\frac{ 2a_{5}}{3}+ 2a_{6} $ \\
			$ \Lambda^{0} K^{+} \bar{K}^{0} $ & $ -\frac{2\sqrt{6}a_{2}}{3}-\frac{ \sqrt{6}a_{3}}{3}- \sqrt{6}a_{4} +\frac{ \sqrt{6}a_{5}}{3} $ \\
			\hline
\end{tabular}
\begin{tabular}{|c|c|}
			\hline
			DCS mode& T-amp/$t_c^2$\\
			\hline
			$ \Sigma^{+} \pi^{0} K^{0} $ & $ -\sqrt{2}a_{3} $ \\
			$ \Sigma^{+} \pi^{-} K^{+} $ & $ 2a_{3} - 2a_{6} $ \\
			$ \Sigma^{+} K^{0} \eta^{0} $ & $ -\frac{\sqrt{6}a_{3}}{3}-\frac{ 2\sqrt{6}a_{4}}{3} $ \\
			$ \Sigma^{0} \pi^{0} K^{+} $ & $ a_{3} - 2a_{6} $ \\
			$ \Sigma^{0} \pi^{+} K^{0} $ & $ \sqrt{2}a_{3} $ \\
			$ \Sigma^{0} K^{+} \eta^{0} $ & $ -\frac{\sqrt{3}a_{3}}{3}-\frac{ 2\sqrt{3}a_{4}}{3} $ \\
			$ \Sigma^{-} \pi^{+} K^{+} $ & $ -2a_{6} $ \\
			$ \Xi^{0} K^{0} K^{+} $ & $ -2a_{4} - 2a_{6} $ \\
			$ \Xi^{-} K^{+} K^{+} $ & $ -4a_{4} - 4a_{6} $ \\
			$ p \pi^{0} \pi^{0} $ & $ 4a_{1} - 2a_{5} $ \\
			$ p \pi^{0} \eta^{0} $ & $ -\frac{2\sqrt{3}a_{5}}{3} $ \\
			$ p \pi^{+} \pi^{-} $ & $ 4a_{1} - 2a_{5} $ \\
			$ p K^{0} \bar{K}^{0} $ & $ 4a_{1} + 2a_{2} + 2a_{3} $ \\
			$ p K^{+} K^{-} $ & $ 4a_{1} + 2a_{2} + 2a_{3} - 2a_{5} - 2a_{6} $ \\
			$ p \eta^{0} \eta^{0} $ & $ 4a_{1} +\frac{ 8a_{2}}{3}+\frac{ 8a_{3}}{3}+\frac{ 8a_{4}}{3}-\frac{ 2a_{5}}{3} $ \\
			$ n \pi^{+} \eta^{0} $ & $ -\frac{2\sqrt{6}a_{5}}{3} $ \\
			$ n K^{+} \bar{K}^{0} $ & $ -2a_{5} - 2a_{6} $ \\
			$ \Lambda^{0} \pi^{0} K^{+} $ & $ \frac{2\sqrt{3}a_{2}}{3}+\frac{ \sqrt{3}a_{3}}{3}+\frac{ 2\sqrt{3}a_{5}}{3} $ \\
			$ \Lambda^{0} \pi^{+} K^{0} $ & $ \frac{2\sqrt{6}a_{2}}{3}+\frac{ \sqrt{6}a_{3}}{3}+\frac{ 2\sqrt{6}a_{5}}{3} $ \\
			
			\hline
\end{tabular}
}
\end{table}

\begin{table}
\caption{T-amplitudes of $ \Xi_{c}^{0}  \to {\bf B_n}MM'$.}\label{Tamp3}
{
\scriptsize
\begin{tabular}{|c|c|}
			\hline
			CF mode& T-amp\\
			\hline
			$ \Sigma^{+} \pi^{0} K^{-} $ & $ \sqrt{2}a_{5} $ \\
			$ \Sigma^{+} \pi^{-} \bar{K}^{0} $ & $ 2a_{5} + 2a_{6} $ \\
			$ \Sigma^{+} K^{-} \eta^{0} $ & $ -\frac{\sqrt{6}a_{5}}{3} $ \\
			$ \Sigma^{0} \pi^{0} \bar{K}^{0} $ & $ a_{2} + a_{4} + a_{5} + 2a_{6} $ \\
			$ \Sigma^{0} \pi^{+} K^{-} $ & $ -\sqrt{2}a_{2} - \sqrt{2}a_{5} $ \\
			$ \Sigma^{0} \bar{K}^{0} \eta^{0} $ & $ \frac{\sqrt{3}a_{2}}{3}-\frac{ \sqrt{3}a_{4}}{3}+\frac{ \sqrt{3}a_{5}}{3} $ \\
			$ \Sigma^{-} \pi^{+} \bar{K}^{0} $ & $ 2a_{4} + 2a_{6} $ \\
			$ \Xi^{0} \pi^{0} \eta^{0} $ & $ \frac{2\sqrt{3}a_{2}}{3}+\frac{ 2\sqrt{3}a_{3}}{3}+\frac{ 2\sqrt{3}a_{4}}{3} $ \\
			$ \Xi^{0} \pi^{+} \pi^{-} $ & $ -4a_{1} - 2a_{2} - 2a_{3} $ \\
$ \Xi^{0} K^{0} \bar{K}^{0} $ & 
$-2(2a_{1} +a_{2}+a_{3}$\\
&$-a_{5}-a_{6})$ \\
			$ \Xi^{0} K^{+} K^{-} $ & $ -4a_{1} + 2a_{5} $ \\
$ \Xi^{0} \eta^{0} \eta^{0} $ & 
$-2(2a_{1}+\frac{a_{2}}{3}+\frac{a_{3}}{3}$\\
&$+\frac{a_{4}}{3}-\frac{4a_{5}}{3})$ \\
			$ \Xi^{-} \pi^{0} \pi^{+} $ & $ \sqrt{2}a_{4} $ \\
			$ \Xi^{-} \pi^{+} \eta^{0} $ & $ -\frac{2\sqrt{6}a_{3}}{3}-\frac{ \sqrt{6}a_{4}}{3} $ \\
			$ \Xi^{-} K^{+} \bar{K}^{0} $ & $ -2a_{3} + 2a_{6} $ \\
			$ p K^{-} \bar{K}^{0} $ & $ 2a_{6} $ \\
			$ n \bar{K}^{0} \bar{K}^{0} $ & $ 4a_{4} + 4a_{6} $ \\
$ \Lambda^{0} \pi^{0} \bar{K}^{0} $ & 
$-\sqrt{3}(\frac{a_{2}}{3}+\frac{ 2a_{3}}{3}+a_{4}+\frac{a_{5}}{3})$ \\
			$ \Lambda^{0} \pi^{+} K^{-} $ & $ \frac{\sqrt{6}a_{2}}{3}+\frac{ 2\sqrt{6}a_{3}}{3}+\frac{ \sqrt{6}a_{5}}{3} $ \\
&\\
&\\
&\\
&\\
&\\
&\\
&\\
&\\
\hline
\end{tabular}
\begin{tabular}{|c|c|}
\hline
CS mode& T-amp/$t_c$ \\
	\hline
	$ \Sigma^{+} \pi^{0} \pi^{-} $ & $ -\sqrt{2}a_{6} $ \\
	$ \Sigma^{+} \pi^{-} \eta^{0} $ & $ \frac{2\sqrt{6}a_{5}}{3}+ \sqrt{6}a_{6} $ \\
	$ \Sigma^{+} K^{0} K^{-} $ & $ 2a_{5} $ \\
	$ \Sigma^{0} \pi^{0} \pi^{0} $ & $ 2\sqrt{2}a_{1} + \sqrt{2}a_{3} - \sqrt{2}a_{5} - 2\sqrt{2}a_{6} $ \\
	$ \Sigma^{0} \pi^{0} \eta^{0} $ & $ -\frac{\sqrt{6}a_{3}}{3}+\frac{ \sqrt{6}a_{4}}{3}+\frac{ \sqrt{6}a_{5}}{3}+ \sqrt{6}a_{6} $ \\
	$ \Sigma^{0} \pi^{+} \pi^{-} $ & $ 2\sqrt{2}a_{1} + \sqrt{2}a_{3} - \sqrt{2}a_{5} $ \\
	$ \Sigma^{0} K^{0} \bar{K}^{0} $ & 
$ \sqrt{2}(2a_{1} +a_{2} +a_{3} +a_{4} -a_{5})$ \\
	$ \Sigma^{0} K^{+} K^{-} $ & $ 2\sqrt{2}a_{1} + \sqrt{2}a_{2} $ \\
$ \Sigma^{0} \eta^{0} \eta^{0} $ & 
$\sqrt{2}(2a_{1} +\frac{ 4a_{2}}{3}+\frac{a_{3}}{3}-\frac{ 2a_{4}}{3}-\frac{a_{5}}{3})$ \\
	$ \Sigma^{-} \pi^{0} \pi^{+} $ & $ -\sqrt{2}a_{6} $ \\
	$ \Sigma^{-} \pi^{+} \eta^{0} $ & $ -\frac{2\sqrt{6}a_{3}}{3}+\frac{ 2\sqrt{6}a_{4}}{3}+ \sqrt{6}a_{6} $ \\
	$ \Sigma^{-} K^{+} \bar{K}^{0} $ & $ -2a_{3} - 2a_{4} $ \\
	$ \Xi^{0} \pi^{-} K^{+} $ & $ 2a_{2} + 2a_{3} + 2a_{5} $ \\
$ \Xi^{0} K^{0} \eta^{0} $ & 
$\sqrt{6}(-\frac{a_{2}}{3}-\frac{a_{3}}{3}
+\frac{ 2a_{4}}{3}-\frac{a_{5}}{3}+a_{6})$ \\
	$ \Xi^{-} \pi^{0} K^{+} $ & $ \sqrt{2}a_{3} - \sqrt{2}a_{4} - \sqrt{2}a_{6} $ \\
	$ \Xi^{-} \pi^{+} K^{0} $ & $ 2a_{3} + 2a_{4} $ \\
	$ p \pi^{0} K^{-} $ & $ -\sqrt{2}a_{5} - \sqrt{2}a_{6} $ \\
	$ p \pi^{-} \bar{K}^{0} $ & $ -2a_{5} $ \\
	$ p K^{-} \eta^{0} $ & $ \frac{\sqrt{6}a_{5}}{3}+ \sqrt{6}a_{6} $ \\
	$ n \pi^{0} \bar{K}^{0} $ & $ \sqrt{2}a_{2} + \sqrt{2}a_{3} + \sqrt{2}a_{5} - \sqrt{2}a_{6} $ \\
	$ n \pi^{+} K^{-} $ & $ -2a_{2} - 2a_{3} - 2a_{5} $ \\
$ n \bar{K}^{0} \eta^{0} $ & 
$\sqrt{6}(\frac{a_{2}}{3}+\frac{a_{3}}{3}+\frac{ 4a_{4}}{3}+\frac{a_{5}}{3}+a_{6})$ \\
$ \Lambda^{0} \pi^{0} \pi^{0} $ & 
$\sqrt{6}(-2a_{1} -\frac{ 2a_{2}}{3}-\frac{a_{3}}{3}+\frac{a_{5}}{3})$ \\
$ \Lambda^{0} \pi^{0} \eta^{0} $ & 
$\sqrt{2}(\frac{2a_{2}}{3}+\frac{a_{3}}{3}-a_{4} -\frac{a_{5}}{3}-a_{6})$ \\
$ \Lambda^{0} \pi^{+} \pi^{-} $ & 
$\sqrt{6}(-2a_{1} -\frac{ 2a_{2}}{3}-\frac{a_{3}}{3}+\frac{a_{5}}{3})$ \\
$ \Lambda^{0} K^{0} \bar{K}^{0} $ & 
$\sqrt{6}(-2a_{1} - a_{2} - a_{3} - a_{4} +a_{5})$ \\
$ \Lambda^{0} K^{+} K^{-} $ & 
$\sqrt{6}(-2a_{1} -\frac{a_{2}}{3}-\frac{ 2a_{3}}{3}+\frac{ 2a_{5}}{3})$ \\
$ \Lambda^{0} \eta^{0} \eta^{0} $ & 
$\sqrt{6}(-2a_{1} -\frac{ 2a_{2}}{3}-a_{3}+\frac{ 2a_{4}}{3}$\\
&$+a_{5} + 2a_{6})$ \\
	\hline
\end{tabular}
\begin{tabular}{|c|c|}
\hline
DCS mode& T-amp/$t_c^2$\\
\hline
$ \Sigma^{+} \pi^{-} K^{0} $ & $ -2a_{6} $ \\
$ \Sigma^{0} \pi^{0} K^{0} $ & $ a_{3} - 2a_{6} $ \\
$ \Sigma^{0} \pi^{-} K^{+} $ & $ -\sqrt{2}a_{3} $ \\
$ \Sigma^{0} K^{0} \eta^{0} $ & $ \frac{\sqrt{3}a_{3}}{3}+\frac{ 2\sqrt{3}a_{4}}{3} $ \\
$ \Sigma^{-} \pi^{0} K^{+} $ & $ \sqrt{2}a_{3} $ \\
$ \Sigma^{-} \pi^{+} K^{0} $ & $ 2a_{3} - 2a_{6} $ \\
$ \Sigma^{-} K^{+} \eta^{0} $ & $ -\frac{\sqrt{6}a_{3}}{3}-\frac{ 2\sqrt{6}a_{4}}{3} $ \\
$ \Xi^{0} K^{0} K^{0} $ & $ -4a_{4} - 4a_{6} $ \\
$ \Xi^{-} K^{0} K^{+} $ & $ -2a_{4} - 2a_{6} $ \\
$ p \pi^{-} \eta^{0} $ & $ -\frac{2\sqrt{6}a_{5}}{3} $ \\
$ p K^{0} K^{-} $ & $ -2a_{5} - 2a_{6} $ \\
$ n \pi^{0} \pi^{0} $ & $ 4a_{1} - 2a_{5} $ \\
$ n \pi^{0} \eta^{0} $ & $ \frac{2\sqrt{3}a_{5}}{3} $ \\
$ n \pi^{+} \pi^{-} $ & $ 4a_{1} - 2a_{5} $ \\
$ n K^{0} \bar{K}^{0} $ & 
$2(2a_{1} + a_{2} + a_{3}$\\
&$ - a_{5} - a_{6})$ \\
			$ n K^{+} K^{-} $ & $ 4a_{1} + 2a_{2} + 2a_{3} $ \\
$ n \eta^{0} \eta^{0} $ & 
$ 4a_{1} +\frac{ 8a_{2}}{3}+\frac{ 8a_{3}}{3}$\\
&$+\frac{ 8a_{4}}{3}-\frac{ 2a_{5}}{3} $ \\
$ \Lambda^{0} \pi^{0} K^{0} $ & $-\sqrt{3}(\frac{2a_{2}}{3}+\frac{a_{3}}{3}+\frac{ 2a_{5}}{3})$ \\
$ \Lambda^{0} \pi^{-} K^{+} $ & 
$\sqrt{6}(\frac{2a_{2}}{3}+\frac{a_{3}}{3}+\frac{ 2a_{5}}{3})$ \\
&\\
&\\
&\\
&\\
&\\
&\\
&\\
&\\
\hline
\end{tabular}
}
\end{table}

\section{Numerical results and discussions}
In the numerical analysis, we perform 
the minimum $\chi^2$ fit to examine if the $SU(3)_f$ symmetry
is valid in the ${\bf B_c}\to {\bf B}_n MM'$ decays.
The equation of the $\chi^2$ fit is given by
\begin{eqnarray}
\chi^2=
\sum_{i} \bigg(\frac{{\cal B}^i_{th}-{\cal B}^i_{ex}}{\sigma_{ex}^i}\bigg)^2\,,
\end{eqnarray}
where ${\cal B}_{th}$ as
${\cal B}({\bf B_c}\to {\bf B}_n MM')$ is calculated by the $SU(3)_f$ parameters, 
and ${\cal B}_{ex}$ the experimental value in Table~\ref{exp}, 
with $\sigma$ the experimental error.
With $\sin\theta_c=0.2248$~\cite{pdg}, one obtains
that $t_c=0.2307$ as the input in Eq.~(\ref{nz_entry}).
The $SU(3)_f$ parameters are written as
\begin{eqnarray}\label{8p}
&&a_1, a_2e^{i\delta_{a_2}},a_3e^{i\delta_{a_3}},
a_4e^{i\delta_{a_4}}, a_5 e^{i\delta_{a_5}},a_6e^{i\delta_{a_6}}\,,
\end{eqnarray}
where the phases $\delta_{a_{2,3,...,6}}$
are due to the nature of  complex numbers associated with  $a_i$,
while $a_1$ can be relatively real. 
This leads to the reduced 11 parameters to be extracted 
with 14 data inputs in Table~\ref{exp}, where the fitting values of $a_i$  and $\delta_{a_{i}}$
are shown in Table~\ref{su3_fit}.
We find that  $\chi^2/d.o.f=8.4/3=2.8$ with $d.o.f$ representing the degree of freedom, 
and we reproduce the branching ratios in the third column of Table~\ref{exp} in order to be compared to the data.
Note that  in calculating the decay branching ratios, we have treated our $SU(3)_f$ parameters as independent ones, which may result in 
 overestimated error ranges in our results.

\begin{table}
\caption{The data of ${\cal B}(\Lambda_c^+\to {\bf B_n}MM)$ 
from the PDG~\cite{pdg}, except for ${\cal B}(\Lambda_c^+\to 
\Sigma^+\pi^0\pi^0,pK^+\pi^-)$~\cite{Berger:2018pli,Aaij:2017rin}.}\label{exp}
\begin{tabular}{|c|c|c|}
\hline
&data&our results\\
\hline
$10^2{\cal B}(\Lambda_c^+\to pK^-\pi^+)$
&$3.4\pm0.4$&$3.3\pm1.0$\\
$10^2{\cal B}(\Lambda_c^+\to p\bar K^0\eta)$
&$1.6\pm0.4$&$0.9\pm0.1$\\
$10^3{\cal B}(\Lambda_c^+\to \Lambda^0 K^+\bar K^0)$
&$5.6\pm1.1$&$5.7\pm1.1$\\
$10^2{\cal B}(\Lambda_c^+\to \Lambda^0 \pi^+\eta)$
&$2.2\pm0.5$&$2.1\pm0.9$\\
$10^2{\cal B}(\Lambda_c^+\to \Sigma^+\pi^+\pi^-)$
&$4.4\pm0.3$&$4.4\pm3.5$\\
$10^2{\cal B}(\Lambda_c^+\to \Sigma^-\pi^+\pi^+)$
&$1.9\pm0.2$&$1.9\pm1.3$\\
$10^2{\cal B}(\Lambda_c^+\to \Sigma^0\pi^+\pi^0)$
&$2.2\pm0.8$&$1.0\pm0.8$\\
$10^2{\cal B}(\Lambda_c^+\to \Sigma^+\pi^0\pi^0)$
&$1.3\pm0.1$&$1.3\pm1.3$\\
$10^3{\cal B}(\Lambda_c^+\to \Sigma^+ K^+ \pi^-)$
&$2.1\pm0.6$&$3.0\pm0.4$\\
\hline
\end{tabular}
\begin{tabular}{|c|c|c|}
\hline
&data&our results\\
\hline
%
%

%
$10^3{\cal B}(\Lambda_c^+\to \Xi^-K^+\pi^+)$
&$6.2\pm0.6$&$6.3\pm0.6$\\ 
$10^2{\cal B}(\Xi_c^+\to \Xi^-\pi^+\pi^+)$
&$6.1\pm 3.1$&$7.2\pm2.0$\\
$10^3{\cal B}(\Lambda_c^+\to p\pi^-\pi^+)$
&$4.2\pm 0.4$&$4.7\pm1.6$\\
$10^4{\cal B}(\Lambda_c^+\to pK^-K^+)$
&$5.2\pm1.2$&$5.1\pm2.1$\\
$10^4{\cal B}(\Lambda_c^+\to pK^+\pi^-)$
&$1.0\pm 0.1$&$1.0\pm0.1$\\
%
&&\\
&&\\
&&\\
&&\\
\hline
\end{tabular}
\end{table}
\begin{table}[h]
	\caption{Fitting results for $a_i$ and $\delta_{a_{i}}$. }\label{su3_fit}
\begin{tabular}{|c|c||c|c|}
	\hline
	$a_i$& result (GeV$^{2}$) &	$\delta_{a_{i}}$&result \\
	\hline
		$a_1$&$9.1\pm 0.6$&--&-- \\
		\hline
	$a_2$&$4.6\pm 0.2$&$\delta_{a_2}$&$164^{\circ}\pm5^{\circ}$ \\
		\hline
	$a_3$&$8.2\pm 0.3$&$\delta_{a_3}$&$135^{\circ}\pm 5^{\circ}$ \\
		\hline
	$a_4$&$2.9\pm 0.4$&$\delta_{a_4}$&$-30^{\circ}\pm 13^{\circ}$\\
		\hline
	$a_5$&$15.4\pm 1.4$&$\delta_{a_5}$&$24^{\circ}\pm 3^{\circ}$ \\
		\hline
	$a_6$&$4.2\pm 0.2$&$\delta_{a_6}$&$120^{\circ}\pm 10^{\circ}$ \\
	\hline
\end{tabular}
\end{table}

To determine the $SU(3)_f$ parameters, 
we use the non-resonant parts of $\Lambda_c^+\to pK^-\pi^+$ 
from the PDG~\cite{pdg}. Note that the resonant 
$\Lambda_c^+\to p(\bar K^{*0}\to)K^-\pi^+$, $K^-(\Delta(1232)^{++}$ $\to)p\pi^+$ and
$\pi^+(\Lambda(1520)\to)pK^-$ contributions are separated
from its total branching ratio.
In addition, the decay of $\Lambda_c^+\to pK^-K^+$ 
is free from the resonant one of $\Lambda_c^+\to p (\phi\to)K^- K^+$.
For the other $\Lambda_c^+$ decays in Table~\ref{exp}, 
some of their resonant parts might be present, but 
taken to be small,
such as 
${\cal B}(\Lambda_c^+\to 
\Sigma^+(\rho^0\to) \pi^+\pi^-)<1.7\%$~\cite{pdg},
which should be insensitive to the fit.
We hence use their total branching ratios,
instead of excluding the resonant contributions.
The $\Xi_c^{0,+}\to {\bf B_n}MM'$ decays are partially observed,
such that we can barely use their data. Nonetheless,
in terms of 
$T(\Lambda_c^+\to \Xi^- K^+\pi^+)=1/(-2t_c) T(\Xi_c^+\to \Sigma^-\pi^+\pi^+)=-2a_6$
and the data of ${\cal B}(\Lambda_c^+\to \Xi^- K^+\pi^+)$, 
we obtain ${\cal B}(\Xi_c^+\to \Sigma^-\pi^+\pi^+)=(1.1\pm 0.1)\times 10^{-2}$ , 
by which the observed ratio of
${\cal B}(\Xi_c^+\to \Sigma^-\pi^+\pi^+)/{\cal B}(\Xi_c^+\to\Xi^-\pi^+\pi^+)=0.18\pm 0.09$ and it
leads to
${\cal B}(\Xi_c^+\to \Xi^-\pi^+\pi^+)=(6.1\pm 3.1)\times 10^{-2}$ as given in Table~\ref{exp}.

With $\chi^2/d.o.f$ being 2.8 in Eq.~(\ref{su3_fit}), it turns out to be a reasonable fit,
so that the $SU(3)_f$ symmetry with the reduced parameters 
can be used to explain the three-body ${\bf B_c}\to {\bf B_n}MM'$ decays.
The relations  of 
$T(\Lambda_c^+\to n \bar K^0\pi^+)=T(\Lambda_c^+\to \Sigma^0 \pi^0\pi^+)$ and
$T(\Lambda_c^+\to \Sigma^0 \pi^0\pi^+)=T(\Lambda_c^+\to \Sigma^- \pi^+\pi^+)/2$
yield
\begin{eqnarray}
&&{\cal B}(\Lambda_c^+\to n \bar K^0\pi^+)\simeq 
{\cal B}(\Lambda_c^+\to \Sigma^0 \pi^0\pi^+)\simeq 
\frac{1}{2}{\cal B}(\Lambda_c^+\to \Sigma^- \pi^+\pi^+)\,,
\end{eqnarray}
which agrees with our numerical analysis. Note that 
the calculation of ${\cal B}(\Lambda_c^+\to \Sigma^- \pi^+\pi^+)$
needs an additional pre-factor of $1/2$ to $T(\Lambda_c^+\to \Sigma^- \pi^+\pi^+)$ 
due to the fact that the $\pi^+\pi^+$ meson-pair involves 
 two identical bosons.

From ${\cal B}(\Lambda_c^+\to p K^+ \pi^-)/{\cal B}(\Lambda_c^+\to p K^- \pi^+)
= \tan^4\theta_c$, 
we find that ${\cal R}_{K\pi}=1.1\pm 0.3$ in the fit without the resonant part. The ratio of  ${\cal R}_{K\pi}\sim 1$ is related to
the same topological diagrams.
 Note that the experimental data of  ${\cal R}_{K\pi}^{Exp}=0.58\pm 0.06$ by LHCb~\cite{Aaij:2017rin} has been obtained 
 by including the resonant contributions in $\Lambda_c^+\to p K^- \pi^+$.
As the prediction from the lowest-wave contributions,
${\cal B}(\Lambda_c^+\to n \pi^{+} \bar{K}^{0},
p\bar K^0 \pi^0)=(0.9\pm 0.8,2.8 \pm 0.6)\times 10^{-2}$
are smaller than the data of 
$(3.6\pm 0.6,4.0\pm0.3)\times 10^{-2}$~\cite{Ablikim:2016mcr,pdg},
which indicate that the resonant and/or high-wave contributions have not been clearly identified  yet.

There exist the sum rules for the T-amplitudes in Table~\ref{Tamp1}.
In particular, by taking the CF $\Lambda_c^+$ decay modes as an example, 
we obtain 
\begin{eqnarray}\label{su3_triangle}
R(\Delta)\equiv T(\Lambda_c^+\to n\bar K^0 \pi^+)-
T(\Lambda_c^+\to pK^- \pi^+)-\sqrt 2 T(\Lambda_c^+\to p\bar K^0 \pi^0)=0\,.\nonumber\\
%
T(\Lambda_c^+\to \Sigma^+\pi^0 \pi^0)-T(\Lambda_c^+\to \Sigma^+\pi^+ \pi^-)
+\frac{1}{2}T(\Lambda_c^+\to \Sigma^-\pi^+ \pi^+)=0\,,\nonumber\\
%
T(\Lambda_c^+\to \Sigma^+ K^0 \bar K^0)-T(\Lambda_c^+\to \Sigma^+ K^+ K^-)
-\sqrt 2 T(\Lambda_c^+\to \Sigma^0 K^+ \bar K^0)=0\,,\nonumber\\
%
T(\Lambda_c^+\to \Xi^0 \pi^+ K^0)-T(\Lambda_c^+\to \Xi^- \pi^+ K^+)
-\sqrt 2 T(\Lambda_c^+\to \Xi^0 \pi^0 K^+)=0\,.
\end{eqnarray}
Note that the first relation of ${\cal R}(\Delta)$ in Eq.~(\ref{su3_triangle}),
which has been used in Ref.~\cite{Ablikim:2016mcr} to reveal the broken isospin symmetry,
is also derived by the isospin symmetry 
in Refs.~\cite{Lu:2016ogy,Gronau:2018vei} with some different signs in the relation 
due to the conventions of the $\pi^{+}$ and  $\bar{K}^{0}$ states.
In addition, the second relation  in Eq.~(\ref{su3_triangle}) can be identified as the special case
in Ref.~\cite{Gronau:2018vei},
 given by
\begin{eqnarray}
T(\Lambda_c^+\to \Sigma^+\pi^0 \pi^0)-T_{sym}(\Lambda_c^+\to \Sigma^+\pi^+ \pi^-)
+\frac{1}{2}T(\Lambda_c^+\to \Sigma^-\pi^+ \pi^+)=0\,,
\end{eqnarray}
with the symmetrized amplitude of 
\begin{eqnarray}
T_{sym}(\Lambda_{c}^+\to\Sigma^+ \pi^+\pi^-)&=&\frac{1}{2}\left[T'(\Lambda_{c}^+\to\Sigma^+ \pi^+ \pi^-)+T'(\Lambda_{c}^+\to\Sigma^+ \pi^- \pi^+)
\right]
\end{eqnarray}
where $T'(\Lambda_{c}^+\to\Sigma^+ \pi^\pm \pi^\mp)$ 
are the amplitudes calculated by the isospin analysis in Ref.~\cite{Gronau:2018vei}.
Likewise, one can take the relations in Eq.~(\ref{su3_triangle}) 
to explore the broken $SU(3)_f$ symmetry.
There are other relations and sum rules obtained from the U-spin symmetry, 
which is also a subgroup of $SU(3)_{f}$~\cite{Grossman:2018ptn}\footnote{There is also a sign issue for the U-spin quantum state 
in Ref.~\cite{Grossman:2018ptn}.}.

The not-yet-observed ${\cal B}(\Lambda_c^+\to {\bf B_n}MM')$ 
can be calculated by the $SU(3)_f$ parameters, 
which are given in Table~\ref{pre_Lc}.
The branching ratios of the three-body $\Xi_c^{+,0}$ decays
are partially observed, such that
we predict ${\cal B}(\Xi_c^{+,0}\to {\bf B_n}MM')$
in Tables~\ref{pre_Xicp} and \ref{pre_Xic0}, respectively,
to be compared to the upcoming data.

\begin{table}[h]
\caption{Numerical results for 
the branching ratios of $\Lambda_{c}^{+} \to {\bf B_n}MM'$,
where ${\cal B}_{{\bf B_n}MM'}\equiv {\cal B}(\Lambda_c^+\to {\bf B_n}MM')$.}\label{pre_Lc}
\begin{tabular}{|c|c|}
\hline
CF mode  & our result\\
\hline
$10^2{\cal B}_{\Sigma^{+} \pi^{0} \eta^{0}}$ & $       3.5 \pm       0.8 $ \\
$10^3{\cal B}_{\Sigma^{+} K^{0} \bar{K}^{0}}$ & $       5.2 \pm       1.2 $ \\
$10^3{\cal B}_{\Sigma^{+} K^{+} {K}^{-}}$ & $       3.0 \pm       0.7 $ \\
$10^7{\cal B}_{\Sigma^{+} \eta^{0} \eta^{0}}$ & $        2.8 \pm        0.6$ \\
$10^2{\cal B}_{\Sigma^{0} \pi^{+} \eta^{0}}$ & $       3.4 \pm       0.8 $ \\
$10^2{\cal B}_{\Sigma^{0} K^{+} \bar{K}^{0}}$ & $       0.5 \pm       0.1 $ \\
$10^2{\cal B}_{\Xi^{0} \pi^{0} K^{+}}$ & $       4.5 \pm       0.8 $ \\
$10^2{\cal B}_{\Xi^{0} \pi^{+} K^{0}}$ & $       8.7 \pm       1.7 $ \\
$10^2{\cal B}_{p \pi^{0} \bar{K}^{0}}$ & $       2.8 \pm       0.6 $ \\
$10^2{\cal B}_{n \pi^{+} \bar{K}^{0}}$ & $       0.9 \pm       0.8 $ \\

&\\
&\\
&\\
&\\
&\\
\hline
\end{tabular}
\begin{tabular}{|c|c|}
\hline
CS mode  & our result\\
\hline
$10^4{\cal B}_{ \Sigma^{+} \pi^{0} K^{0}}$ & $      8.6 \pm       2.6 $ \\

$10^5{\cal B}_{  \Sigma^{+} K^{0} \eta^{0}} $ & $       3.5 \pm       0.4 $ \\
$10^3{\cal B}_{  \Sigma^{0} \pi^{0} K^{+}} $ & $      1.2 \pm       0.3 $ \\
$10^4{\cal B}_{  \Sigma^{0} \pi^{+} K^{0}} $ & $      8.3 \pm       2.5 $ \\
$10^5{\cal B}_{  \Sigma^{0} K^{+} \eta^{0}} $ & $       1.8 \pm       0.2 $ \\
$10^4{\cal B}_{  \Sigma^{-} \pi^{+} K^{+}} $ & $      3.3 \pm       2.3 $ \\
$10^3{\cal B}_{  p \pi^{0} \pi^{0}} $ & $      2.4 \pm       0.8 $ \\
$10^3{\cal B}_{  p \pi^{0} \eta^{0}} $ & $      3.7 \pm       0.9 $ \\
$10^3{\cal B}_{  p k^{0} \bar{K}^{0}} $ & $      4.3 \pm       1.0 $ \\
$10^4{\cal B}_{  p \eta^{0} \eta^{0}} $ & $      4.7 \pm       1.0 $ \\
$10^3{\cal B}_{  n \pi^{+} \eta^{0}} $ & $      7.3 \pm       1.8 $ \\
$10^3{\cal B}_{  n K^{+} \bar{K}^{0}} $ & $      5.9 \pm       1.3 $ \\
$10^3{\cal B}_{  \Lambda^{0} \pi^{0} K^{+}} $ & $       4.5 \pm      0. 8 $ \\
$10^3{\cal B}_{  \Lambda^{0} \pi^{+} K^{0}} $ & $       8.8 \pm       1.5 $ \\
$10^4{\cal B}_{  \Lambda^{0} K^{+} \eta^{0}} $ & $       1.9 \pm       0.6 $ \\

\hline
\end{tabular}
\begin{tabular}{|c|c|}
\hline
DCS mode&our result\\
\hline
$10^6{\cal B}_ {\Sigma^{+} K^{0} K^{0} } $ & $      2.0 \pm       0.5 $ \\
$10^6{\cal B}_ { \Sigma^{0} K^{0} K^{+} } $ & $      2.0 \pm       0.6 $ \\
$10^6{\cal B}_ { \Sigma^{-} K^{+} K^{+} } $ & $      2.0 \pm       0.5 $ \\
$10^5{\cal B}_ { p \pi^{0} K^{0} } $ & $      5.0 \pm       0.5 $ \\

$10^5{\cal B}_ { n \pi^{0} K^{+} } $ & $      5.0 \pm      0.5 $ \\
$10^4{\cal B}_ { n \pi^{+} K^{0} } $ & $     1.0 \pm 0.1 $ \\
&   \\
&   \\
&   \\
&   \\
&   \\
&   \\
&   \\
&   \\	
&   \\														
\hline
\end{tabular}
\end{table}

\begin{table}
\caption{
Numerical results for 
the branching ratios of $\Xi_{c}^{+} \to {\bf B_n}MM'$,
where ${\cal B}_{{\bf B_n}MM'}\equiv {\cal B}(\Xi_{c}^+\to {\bf B_n}MM')$.}\label{pre_Xicp}
\begin{tabular}{|c|c|}
\hline
CF mode& our result\\
\hline
$10^3{\cal B}_{ \Sigma^{+} \pi^{0} \bar{K}^{0}}$ & $       5.4 \pm       4.0 $ \\
$10^2{\cal B}_{ \Sigma^{+} \pi^{+} K^{-}}$ & $       6.1 \pm       0.6$ \\
$10^3{\cal B}_{ \Sigma^{+} \bar{K}^{0} \eta^{0}} $ & $       4.6 \pm       0.6 $ \\
$10^2{\cal B}_{ \Sigma^{0} \pi^{+} \bar{K}^{0}} $ & $       1.2 \pm       0.3 $ \\
$10^2{\cal B}_{ \Xi^{0} \pi^{0} \pi^{+}} $ & $       1.9 \pm       0.5 $ \\
$10^2{\cal B}_{ \Xi^{0} \pi^{+} \eta^{0}} $ & $       1.0 \pm       0.2 $ \\
$10^3{\cal B}_{ \Xi^{0} K^{+} \bar{K}^{0}} $ & $       4.9 \pm       0.5 $ \\
$10^2{\cal B}_{ p \bar{K}^{0} \bar{K}^{0}} $ & $      4.3 \pm       1.2 $ \\
$10^2{\cal B}_{ \Lambda^{0} \pi^{+} \bar{K}^{0} }$ & $      4.6 \pm       1.2 $ \\
&\\
&\\
&\\
&\\
&\\
&\\
&\\
&\\
&\\
&\\
&\\
\hline
\end{tabular}
\begin{tabular}{|c|c|}
\hline
CS mode&our result\\
\hline
$10^3{\cal B}_{  \Sigma^{+} \pi^{0} \eta^{0}  }$ & $      9.6 \pm       1.8 $ \\
$10^3{\cal B}_{  \Sigma^{+} \pi^{+} \pi^{-} } $ & $      5.1 \pm      2.0$ \\
$10^3{\cal B}_{  \Sigma^{+} K^{0} \bar{K}^{0}  }$ & $      5.4 \pm       1.3 $ \\
$10^3{\cal B}_{  \Sigma^{+} K^{+} K^{-}  }$ & $      1.0 \pm       0.4 $ \\
$10^4{\cal B}_{  \Sigma^{+} \eta^{0} \eta^{0}  }$ & $      1.8 \pm       1.0 $ \\
$10^3{\cal B}_{  \Sigma^{0} \pi^{0} \pi^{+}  }$ & $      5.6 \pm       0.5$ \\
$10^3{\cal B}_{  \Sigma^{0} \pi^{+} \eta^{0}  }$ & $      9.4 \pm       1.8 $ \\
$10^3{\cal B}_{  \Sigma^{0} K^{+} \bar{K}^{0}  }$ & $      4.4 \pm       0.9 $ \\
$10^2{\cal B}_{  \Sigma^{-} \pi^{+} \pi^{+}  }$ & $      1.1 \pm       0.1 $ \\
$10^3{\cal B}_{  \Xi^{0} \pi^{0} K^{+} }$  & $      6.4 \pm       1.6 $ \\
$10^2{\cal B}_{  \Xi^{0} \pi^{+} K^{0} }$  & $      1.9 \pm       0.4 $ \\
$10^4{\cal B}_{  \Xi^{0} K^{+} \eta^{0}  }$ & $       1.3 \pm       0.3 $ \\
$10^4{\cal B}_{  \Xi^{-} \pi^{+} K^{+}  }$ & $      8.3 \pm      5.3 $ \\
$10^2{\cal B}_{  p \pi^{0} \bar{K}^{0}  }$ & $     2.4 \pm      0.2 $ \\
$10^2{\cal B}_{  p \pi^{+} K^{-}  }$ & $      2.4 \pm      0.3 $ \\
$10^3{\cal B}_{  n \pi^{+} \bar{K}^{0}  }$ & $      5.5 \pm       0.5 $ \\
$10^2{\cal B}_{  \Lambda^{0} \pi^{+} \eta^{0}  }$ & $      1.7 \pm      0.3 $ \\
$10^3{\cal B}_{  \Lambda^{0} K^{+} \bar{K}^{0}  }$ & $      4.7 \pm      1.0 $ \\
&\\
&\\
\hline
\end{tabular}
\begin{tabular}{|c|c|}
\hline
DCS mode& our result\\
\hline
$10^4{\cal B}_ { \Sigma^{+} \pi^{0} K^{0} }$ & $     2.6 \pm      0.2 $ \\
$10^4{\cal B}_ { \Sigma^{+} \pi^{-} K^{+}  }$ & $     1.4 \pm      0.3 $ \\
$10^6{\cal B}_ { \Sigma^{+} K^{0} \eta^{0}  }$ & $      2.0 \pm      1.4 $ \\
$10^6{\cal B}_ { \Sigma^{0} \pi^{0} K^{+}  }$ & $      7.6 \pm      5.9 $ \\
$10^4{\cal B}_ { \Sigma^{0} \pi^{+} K^{0}  }$ & $     2.5 \pm      0.2 $ \\
$10^6{\cal B}_ { \Sigma^{0} K^{+} \eta^{0} } $ & $      1.0 \pm      0.7 $ \\
$10^4{\cal B}_ { \Sigma^{-} \pi^{+} K^{+}  }$ & $      1.3 \pm       0.1 $ \\
$10^6{\cal B}_ { \Xi^{0} K^{0} K^{+}  }$ & $      3.0 \pm       1.9 $ \\
$10^6{\cal B}_ { \Xi^{-} K^{+} K^{+}  }$ & $      5.7 \pm      3.2 $ \\
$10^4{\cal B}_ { p \pi^{0} \pi^{0}  }$ & $      7.2 \pm      1.8 $ \\
$10^3{\cal B}_ { p \pi^{0} \eta^{0}  }$ & $      1.1 \pm      0.2 $ \\
$10^3{\cal B}_ { p \pi^{+} \pi^{-}  }$ & $      1.4 \pm      0.4 $ \\
$10^4{\cal B}_ { p K^{0} \bar{K}^{0}  }$ & $     7.7 \pm     1.7 $ \\
$10^4{\cal B}_ { p K^{+} K^{-}  }$ & $     1.6 \pm     1.2 $ \\
$10^5{\cal B}_ { p \eta^{0} \eta^{0}  }$ & $      9.3 \pm      4.5 $ \\
$10^3{\cal B}_ { n \pi^{+} \eta^{0}  }$ & $     2.1 \pm      0.4 $ \\
$10^3{\cal B}_ { n K^{+} \bar{K}^{0}  }$ & $     1.6 \pm      0.3 $ \\
$10^4{\cal B}_ { \Lambda^{0} \pi^{0} K^{+}  }$ & $      5.0 \pm      1.0 $ \\
$10^4{\cal B}_ { \Lambda^{0} \pi^{+} K^{0}  }$ & $    9.7 \pm      2.0 $ \\
$10^5{\cal B}_ { \Lambda^{0} K^{+} \eta^{0}  }$ & $     9.0 \pm      2.2 $ \\

\hline
\end{tabular}
\end{table}

\begin{table}
\caption{Numerical results for 
the branching ratios of $\Xi_{c}^{0} \to {\bf B_n}MM'$,
where ${\cal B}_{{\bf B_n}MM'}\equiv {\cal B}(\Xi_{c}^0\to {\bf B_n}MM')$.}\label{pre_Xic0}
\begin{tabular}{|c|c|}
\hline
CF mode& our result\\
\hline
$10^2{\cal B}_{ \Sigma^{+} \pi^{0} K^{-} }$ & $       8.8 \pm       1.5 $ \\
$10^1{\cal B}_{ \Sigma^{+} \pi^{-} \bar{K}^{0}} $ & $       1.8 \pm       0.3 $ \\
$10^3{\cal B}_{ \Sigma^{+} K^{-} \eta^{0}} $ & $       5.2 \pm       0.9 $ \\
$10^2{\cal B}_{ \Sigma^{0} \pi^{0} \bar{K}^{0}} $ & $       4.4 \pm       1.1 $ \\
$10^2{\cal B}_{ \Sigma^{0} \pi^{+} K^{-} }$ & $       5.4 \pm       1.2 $ \\
$10^3{\cal B}_{ \Sigma^{0} \bar{K}^{0} \eta^{0} }$ & $       1.4 \pm       0.3 $ \\
$10^2{\cal B}_{ \Xi^{0} \pi^{0} \pi^{0}} $ & $       8.1 \pm       1.9 $ \\
$10^2{\cal B}_{ \Xi^{0} \pi^{0} \eta^{0}} $ & $       1.2 \pm       0.2 $ \\
$10^1{\cal B}_{ \Xi^{0} \pi^{+} \pi^{-} }$ & $       1.3 \pm       0.3 $ \\
$10^3{\cal B}_{ \Xi^{0} K^{+} K^{-} }$ & $       3.6 \pm       0.9 $ \\
$10^4{\cal B}_{ \Xi^{0} \eta^{0} \eta^{0} }$ & $       2.2 \pm       0.9 $ \\
$10^3{\cal B}_{ \Xi^{-} \pi^{0} \pi^{+} }$ & $       4.6 \pm       1.2 $ \\
$10^2{\cal B}_{ \Xi^{-} \pi^{+} \eta^{0} }$ & $       1.1 \pm       0.1 $ \\
$10^2{\cal B}_{ p K^{-} \bar{K}^{0} }$ & $       1.2 \pm       0.1 $ \\
$10^3{\cal B}_{ n \bar{K}^{0} \bar{K}^{0} }$ & $       6.4 \pm       6.3 $ \\
$10^2{\cal B}_{ \Lambda^{0} \pi^{0} \bar{K}^{0} }$ & $       2.0 \pm       0.6 $ \\
$10^2{\cal B}_{ \Lambda^{0} \pi^{+} K^{-} }$ & $       5.9 \pm       0.8 $ \\
&\\
&\\
&\\
&\\
&\\
&\\
&\\
&\\
\hline
\end{tabular}
\begin{tabular}{|c|c|}
\hline
CS mode& our result\\
\hline
$10^4{\cal B}_{   \Sigma^{+} \pi^{0} \pi^{-}}$ & $       7.2 \pm       0.7 $ \\
$10^3{\cal B}_{   \Sigma^{+} \pi^{-} \eta^{0} }$ & $       5.7 \pm      0.9 $ \\
$10^3{\cal B}_{   \Sigma^{+} K^{0} K^{-} }$ & $       2.4 \pm       0.4 $ \\
$10^3{\cal B}_{   \Sigma^{0} \pi^{0} \pi^{0} }$ & $       1.3 \pm       0.3 $ \\
$10^3{\cal B}_{   \Sigma^{0} \pi^{0} \eta^{0} }$ & $       1.9 \pm       0.4 $ \\
$10^4{\cal B}_{   \Sigma^{0} K^{+} K^{-} }$ & $       9.7 \pm       1.7 $ \\
$10^5{\cal B}_{   \Sigma^{0} \eta^{0} \eta^{0} }$ & $       2.3 \pm       1.2 $ \\
$10^4{\cal B}_{   \Sigma^{-} \pi^{0} \pi^{+} }$ & $       7.1 \pm       0.6 $ \\
$10^4{\cal B}_{   \Sigma^{-} \pi^{+} \eta^{0} }$ & $       6.3 \pm       2.0 $ \\
$10^4{\cal B}_{   \Sigma^{-} K^{+} \bar{K}^{0} }$ & $       2.9 \pm       0.6 $ \\
$10^3{\cal B}_{   \Xi^{0} \pi^{0} K^{0} }$ & $      3.0 \pm       0.7 $ \\
$10^3{\cal B}_{   \Xi^{0} \pi^{-} K^{+} }$ & $      4.8 \pm       0.9 $ \\
$10^4{\cal B}_{   \Xi^{-} \pi^{0} K^{+} }$ & $       6.2 \pm       1.3 $ \\
$10^4{\cal B}_{   \Xi^{-} \pi^{+} K^{0} }$ & $      7.2 \pm       1.5 $ \\
$10^3{\cal B}_{   p \pi^{0} K^{-} }$ & $      9.5 \pm       1.6 $ \\
$10^2{\cal B}_{   p \pi^{-} \bar{K}^{0} }$ & $      1.9 \pm       0.3 $ \\
$10^3{\cal B}_{   p K^{-} \eta^{0} }$ & $       1.8 \pm       0.3 $ \\
$10^3{\cal B}_{   n \pi^{0} \bar{K}^{0} }$ & $       5.2 \pm       1.3 $ \\
$10^2{\cal B}_{   n \pi^{+} K^{-} }$ & $      1.5 \pm       0.3 $ \\
$10^3{\cal B}_{   n \bar{K}^{0} \eta^{0} }$ & $       1.9 \pm       0.6 $ \\
$10^3{\cal B}_{   \Lambda^{0} \pi^{0} \pi^{0} }$ & $      5.3 \pm       1.5 $ \\
$10^3{\cal B}_{   \Lambda^{0} \pi^{0} \eta^{0} }$ & $      2.2 \pm       0.4 $ \\
$10^2{\cal B}_{   \Lambda^{0} \pi^{+} \pi^{-} }$ & $      1.1 \pm       0.3 $ \\
$10^4{\cal B}_{  \Lambda^{0} K^{+} K^{-} }$ & $       3.0 \pm       2.5 $ \\
$10^4{\cal B}_{   \Lambda^{0} \eta^{0} \eta^{0} }$ & $       2.4\pm       1.4 $ \\
\hline
\end{tabular}
\begin{tabular}{|c|c|}
\hline
DCS mode& our result\\
\hline
$10^5{\cal B}_ { \Sigma^{+} \pi^{-} K^{0}}$ & $      3.4 \pm       0.3 $ \\
$10^5{\cal B}_ { \Sigma^{0} \pi^{-} K^{+} }$ & $      6.5 \pm       0.5 $ \\
$10^7{\cal B}_ { \Sigma^{0} K^{0} \eta^{0} }$ & $       2.6 \pm       1.7 $ \\
$10^5{\cal B}_ { \Sigma^{-} \pi^{0} K^{+} }$ & $      6.4 \pm       0.5 $ \\
$10^5{\cal B}_ { \Sigma^{-} \pi^{+} K^{0} }$ & $      3.4 \pm      0.7 $ \\
$10^7{\cal B}_ { \Sigma^{-} K^{+} \eta^{0} }$ & $      5.1 \pm       3.4 $ \\
$10^6{\cal B}_ { \Xi^{0} K^{0} K^{0} }$ & $       1.5 \pm       1.1 $ \\
$10^7{\cal B}_ { \Xi^{-} K^{0} K^{+} }$ & $       7.1 \pm       6.7 $ \\
$10^4{\cal B}_ { p \pi^{-} \eta^{0} }$ & $      5.4 \pm      0.9 $ \\
$10^4{\cal B}_ { p K^{0} K^{-} }$ & $      4.2 \pm      0.7 $ \\
$10^4{\cal B}_ { n \pi^{0} \pi^{0} }$ & $      1.8 \pm       0.5 $ \\
$10^4{\cal B}_ { n \pi^{0} \eta^{0} }$ & $      2.7 \pm       0.5 $ \\
$10^4{\cal B}_ { n \pi^{+} \pi^{-} }$ & $      3.6 \pm       0.9 $ \\
$10^5{\cal B}_ { n K^{0} \bar{K}^{0} }$ & $      3.9 \pm      2.9 $ \\
$10^4{\cal B}_ { n K^{+} K^{-} }$ & $     2.0 \pm      0.5 $ \\
$10^5{\cal B}_ { n \eta^{0}\eta^{0} }$ & $     2.4 \pm      1.2 $ \\
$10^4{\cal B}_ { \Lambda^{0} \pi^{0} K^{0} }$ & $      1.3 \pm      0.3 $ \\
$10^4{\cal B}_ { \Lambda^{0} \pi^{-} K^{+} }$ & $      2.5 \pm      0.5 $ \\
$10^5{\cal B}_ { \Lambda^{0}K^{0} \eta^{0} }$ & $      2.3 \pm      0.6 $ \\
&\\
&\\
&\\
&\\
&\\
&\\
\hline
\end{tabular}
\end{table}

\section{Conclusions}
We have studied the three-body anti-triplet ${\bf B_c}\to {\bf B_n}MM'$ decays
in the approach of the $SU(3)_f$ symmetry.
In our analysis, we have only concentrated on the S-wave $MM'$-pair contributions, 
so that the decays of ${\bf B_c}\to {\bf B_n}MM'$ can be decomposed into irreducible forms with 11 parameters under $SU(3)_f$.
With the minimum $\chi^2$ fit to the 14 existing data points,  we have obtained a reasonable value of $\chi^2/d.o.f=2.8$.
With our numerical results, we have shown the same triangle relation of 
${\cal A}(\Lambda_c^+\to n\bar K^0 \pi^+)-
{\cal A}(\Lambda_c^+\to pK^- \pi^+)-\sqrt 2 {\cal A}(\Lambda_c^+\to p\bar K^0 \pi^0)=0$ under $SU(3)_f$ as that based on the isospin symmetry.
In addition,  
 for the CF decays, we have obtained the  sum rules of
 ${\cal A}(\Lambda_c^+\to \Sigma^+\pi^0 \pi^0)-{\cal A}(\Lambda_c^+\to \Sigma^+\pi^+ \pi^-)
+1/2{\cal A}(\Lambda_c^+\to \Sigma^-\pi^+ \pi^+)=0$,
${\cal A}(\Lambda_c^+\to \Sigma^+ K^0 \bar K^0)-{\cal A}(\Lambda_c^+\to \Sigma^+ K^+ K^-)
-\sqrt 2 {\cal A}(\Lambda_c^+\to \Sigma^0 K^+ \bar K^0)=0$
and
${\cal A}(\Lambda_c^+\to \Xi^0 \pi^+ K^0)-{\cal A}(\Lambda_c^+\to \Xi^- \pi^+ K^+)
-\sqrt 2 {\cal A}(\Lambda_c^+\to \Xi^0 \pi^0 K^+)=0$.
Furthermore,
we have predicted that 
${\cal B}(\Lambda_c^+\to n \pi^{+} \bar{K}^{0})=(0.9\pm 0.8)\times 10^{-2}$,
which is (3-4) times smaller than the BESIII observation of $(3.6\pm 0.6)\times 10^{-2}$.
This indicates that there are some  contributions from the resonant and/or P-wave  states.
For the to-be-observed $\Lambda_c^+\to {\bf B_n}MM'$ and
the partial observed $\Xi_c^{0,+}\to {\bf B_n}MM'$ decays,
the branching ratios have been calculated with the $SU(3)_f$ amplitudes,
to be compared to the future measurements by  BESIII and LHCb.

\section*{ACKNOWLEDGMENTS}
This work was supported in part by National Center for Theoretical Sciences,
MoST (MoST-104-2112-M-007-003-MY3 and MoST-107-2119-M-007-013-MY3), and
National Science Foundation of China (11675030).



\begin{thebibliography}{99}

\bibitem{Zupanc:2013iki} 
A.~Zupanc {\it et al.} [Belle Collaboration],
Phys.\ Rev.\ Lett.\  {\bf 113}, 042002 (2014). 

\bibitem{Ablikim:2015flg} 
M.~Ablikim {\it et al.} [BESIII Collaboration],
Phys.\ Rev.\ Lett.\  {\bf 116}, 052001 (2016). 

\bibitem{Berger:2018pli} 
M.~Berger {\it et al.} [Belle Collaboration],
arXiv:1802.03421 [hep-ex].

\bibitem{Ablikim:2016mcr} 
M.~Ablikim {\it et al.} [BESIII Collaboration],
Phys.\ Rev.\ Lett.\  {\bf 118}, 112001 (2017). 

\bibitem{pdg}
C.~Patrignani {\it et al.} [Particle Data Group],
Chin.\ Phys.\ C {\bf 40}, 100001 (2016).

\bibitem{Lu:2016ogy} 
C.D.~Lu, W.~Wang and F.S.~Yu,
Phys.\ Rev.\ D {\bf 93}, 056008 (2016). 

\bibitem{Gronau:2018vei} 

M.~Gronau, J.~L.~Rosner and C.~G.~Wohl,
Phys.\ Rev.\ D {\bf 97}, no. 11, 116015 (2018);
Addendum: [Phys.\ Rev.\ D {\bf 98}, no. 7, 073003 (2018)].



\bibitem{Grossman:2018ptn} 
Y.~Grossman and S.~Schacht,
Phys.\ Rev.\ D {\bf 99}, 033005 (2019).

\bibitem{Yang:2015ytm} 
S.B.~Yang {\it et al.} [Belle Collaboration],
Phys.\ Rev.\ Lett.\  {\bf 117}, 011801 (2016). 

\bibitem{Aaij:2017rin} 
R.~Aaij {\it et al.} [LHCb Collaboration],
JHEP {\bf 1803}, 043 (2018). 

\bibitem{Bjorken:1988ya} 
J.D.~Bjorken, 
Phys.\ Rev.\ D {\bf 40}, 1513 (1989).

\bibitem{ali} A. Ali, G. Kramer and C.D. Lu, Phys. Rev.  D{\bf 58}, 094009 (1998).

\bibitem{Geng:2006jt} 
C.Q.~Geng, Y.K.~Hsiao and J.N.~Ng,
Phys.\ Rev.\ Lett.\  {\bf 98}, 011801 (2007).

\bibitem{Hsiao:2014mua} 
Y.K.~Hsiao and C.Q.~Geng,
Phys.\ Rev.\ D {\bf 91}, 116007 (2015).

\bibitem{Cheng:1991sn}
H.Y.~Cheng and B.~Tseng,
Phys.\ Rev.\ D {\bf 46}, 1042 (1992); {\bf 55}, 1697(E) (1997).

\bibitem{Cheng:1993gf}
H.Y.~Cheng and B.~Tseng,
Phys.\ Rev.\ D {\bf 48}, 4188 (1993).

\bibitem{Zenczykowski:1993hw}
P.~Zenczykowski,
Phys.\ Rev.\ D {\bf 50}, 402 (1994).


\bibitem{Fayyazuddin:1996iy}
Fayyazuddin and Riazuddin,
Phys.\ Rev.\ D {\bf 55}, 255; {\bf 56}, 531(E) (1997).


\bibitem{Dhir:2015tja}
R.~Dhir and C.S.~Kim,
Phys.\ Rev.\ D {\bf 91}, 114008 (2015). 

\bibitem{Cheng:2018hwl}
H.Y.~Cheng, X.W.~Kang and F.~Xu,
Phys.\ Rev.\ D {\bf 97}, 074028 (2018).
 



\bibitem{He:2000ys}
X.G.~He, Y.K.~Hsiao, J.Q.~Shi, Y.L.~Wu and Y.F.~Zhou,
Phys.\ Rev.\ D {\bf 64}, 034002 (2001).

\bibitem{Fu:2003fy}
H.K.~Fu, X.G.~He and Y.K.~Hsiao,
Phys.\ Rev.\ D {\bf 69}, 074002 (2004).

\bibitem{Hsiao:2015iiu}
Y.K.~Hsiao, C.F.~Chang and X.G.~He,
Phys.\ Rev.\ D {\bf 93}, 114002 (2016).

\bibitem{He:2015fwa}
X.G.~He and G.N.~Li,
Phys.\ Lett.\ B {\bf 750}, 82 (2015).

\bibitem{He:2015fsa}
M.~He, X.G.~He and G.N.~Li,
Phys.\ Rev.\ D {\bf 92}, 036010 (2015).

\bibitem{Grossman:2012ry}
Y.~Grossman and D.J.~Robinson, 
JHEP {\bf 1304}, 067 (2013). 

\bibitem{Pirtskhalava:2011va}
D.~Pirtskhalava and P.~Uttayarat,
Phys.\ Lett.\ B {\bf 712}, 81 (2012). 

\bibitem{Cheng:2012xb}
H.Y.~Cheng and C.W.~Chiang,
Phys.\ Rev.\ D {\bf 86}, 014014 (2012). 

\bibitem{Savage:1989qr}
M.J.~Savage and R.P.~Springer,
Phys.\ Rev.\ D {\bf 42}, 1527 (1990).

\bibitem{Savage:1991wu}
M.J.~Savage,
Phys.\ Lett.\ B {\bf 257}, 414 (1991).

\bibitem{h_term}
G.~Altarelli, N.~Cabibbo and L.~Maiani,
Phys.\ Lett.\  {\bf 57B}, 277 (1975).


\bibitem{Geng:2017esc}
C.Q.~Geng, Y.K.~Hsiao, Y.H.~Lin and L.L. Liu,
Phys.\ Lett.\ B {\bf 776}, 265 (2017). 

\bibitem{Geng:2018plk}
C.Q.~Geng, Y.K.~Hsiao, C.W.~Liu and T.H.~Tsai,
Phys.\ Rev.\ D {\bf 97}, 073006 (2018).

\bibitem{Geng:2017mxn}
C.Q.~Geng, Y.K.~Hsiao, C.W.~Liu and T.H.~Tsai,
JHEP {\bf 1711}, 147 (2017). 

\bibitem{Geng:2018bow} 
C.Q.~Geng, Y.K.~Hsiao, C.W.~Liu and T.H.~Tsai,
Eur.\ Phys.\ J.\ C {\bf 78}, 593 (2018). 
\bibitem{Wang:2017azm}
W.~Wang, Z.P.~Xing and J.~Xu,
Eur.\ Phys.\ J.\ C {\bf 77}, 800 (2017). 

\bibitem{Wang:2017gxe}
D.~Wang, P.F.~Guo, W.H.~Long and F.S.~Yu,
JHEP {\bf 1803}, 066 (2018).

\bibitem{Geng:2018rse} 
  C.Q.~Geng, C.W.~Liu and T.H.~Tsai,
  Phys.\ Lett.\ B {\bf 790}, 225 (2019).
  
\bibitem{Buras:1998raa} 
A.J.~Buras, hep-ph/9806471.

\bibitem{Li:2012cfa}
H.n.~Li, C.D.~Lu and F.S.~Yu,
Phys.\ Rev.\ D {\bf 86}, 036012 (2012).

\bibitem{Fajfer:2002gp}
S.~Fajfer, P.~Singer and J.~Zupan,
Eur.\ Phys.\ J.\ C {\bf 27}, 201 (2003).

\end{thebibliography}
\end{document}